\documentclass[12pt]{iopart}

\usepackage{amssymb}
\usepackage{color}
\usepackage{graphicx}
\usepackage{subfigure}

\begin{document}
\def\bbm[#1]{\mbox{\boldmath$#1$}}

\title[Non-Markovian dynamics: time-independent parameters]{A tomographic approach to non-Markovian master equations}
\author{Bruno Bellomo$^{1,2}$, Antonella De Pasquale$^{1,3}$, Giulia Gualdi$^{1,4}$ and Ugo Marzolino$^{1,5}$}
\address{$^1$ MECENAS, Universit\`a Federico II di Napoli \& Universit\`a di Bari, Italy}
\address{$^2$  CNISM and Dipartimento di Scienze Fisiche ed Astronomiche, Universit\`a di  Palermo, via Archirafi 36, 90123 Palermo, Italy}
\address{$^3$ Dipartimento di Fisica, Universit\`{a} di Bari, via Amendola 173, I-70126 Bari, Italy; \\ INFN, Sezione di Bari, I-70126 Bari, Italy}
\address{$^4$  Dipartimento di Matematica e Informatica, Universit\`a degli Studi di Salerno, Via Ponte don Melillo, I-84084 Fisciano (SA), Italy\\ CNR-INFM Coherentia, Napoli, Italy; CNISM Unit\`a di Salerno; INFN Sezione di Napoli gruppo collegato di Salerno, Italy}
\address{$^5$ Dipartimento di Fisica, Universit\`{a} di Trieste, Strada Costiera 11, 34151, Trieste, Italy;\\ INFN, Sezione di Trieste, 34151, Trieste, Italy}

\begin{abstract}
We  propose a procedure based on symplectic tomography for reconstructing the unknown parameters of a convolutionless non-Markovian Gaussian noisy evolution. Whenever the time-dependent master equation coefficients  are given as a function of some unknown time-independent parameters, we show that these parameters can be reconstructed by means of a finite number of tomograms. Two different approaches towards reconstruction, integral and differential,  are presented and applied to a benchmark model made of a harmonic oscillator coupled to a bosonic bath. For this model the number of tomograms needed to retrieve  the  unknown parameters  is explicitly computed. 

\end{abstract}


\maketitle

\section{Introduction}
A central issue in modern physics is the investigation of  noise as it drastically affects the evolution of quantum systems. Such a general phenomenon is of importance in quantum information science and beyond, as it addresses a fundamental issue in quantum theory \cite{Petruccione-Breuerlibro2002, Gardiner-Zoller}.
So far, two main dynamical regimes, Markovian and non-Markovian, can usually be distinguished according to the noise time scale (respectively shorter or longer than that of system dynamics).
Here we address the non-Markovian case. In facts, despite Markovian evolutions  have been exstensively investigated,  in general real noisy dynamics are far from being Markovian. Although gaining much interest in the last years both in theory and experiment \cite{Hope2000,Pomyalov2005}, a general theory for non-Markovian dynamics is still missing. In this dynamical regime,  exact master equations have been derived for the evolution of a Brownian particle linearly coupled to a harmonic oscillator bath, for instance via path integral methods \cite{HPZ1992,HM1994} or phase-space and Wigner function computations \cite{HR1985,HY1996}.  Analogous results are derived by means of quantum trajectories, either exactly or in  weak coupling approximation \cite{Diosi1997,Strunz2004,Yu2004,Bassi2009}. In the framework of path integral methods, master equations have been derived both for initially correlated states \cite{KG1997,RP1997}, and for factorized initial states in the case of weak non linear interactions \cite{HPZ1993}. In general, all these master equations cover only few cases and are not simple to solve. Indeed, it would be highly desirable to find an approximation scheme fully capturing non-Markovian features.
In general, different approximations (for example on system-bath interaction strength) may  lead to irreconcilable dynamics \cite{Petruccione-Breuerlibro2002}.\\
For Markovian dissipation it has been shown that, by exploiting symplectic tomography \cite{Lvovsky2009} and using Gaussian probes, unknown parameters governing the dynamics of the system can be reconstructed trough a limited number of tomographic measurements \cite{Bellomo2009}. In this paper we extend this tomography-based approach  to time-dependent dissipative dynamics. We present an experimentally feasible procedure that exploits symplectic tomography and which, under the assumption of Gaussian noise, allows  to reconstruct unknown time-independent parameters that characterize the time-dependent coefficients of the master equation. Even though the assumption of Gaussian noise might be seen as an idealization, it is actually well fitted for a significant number of models \cite{Petruccione-Breuerlibro2002,Gardiner-Zoller}. Also, small deviations from Gaussianity would introduce small and controllable errors. We also note that Gaussian probes are quite straightforward to produce \cite{Mandel1995}.\\
The paper is organized as follows.  In section \ref{par:tomogramscumulants procedure}
we briefly recall the procedure that allows to reconstruct cumulants of a Gaussian Êstate (our probe) undergoing a dissipative dynamics.
In section \ref{par:Description of the system} we introduce the class of non-Markovian master equations we will study, and derive the expressions for the first and second time-dependent momenta (cumulants). In section \ref{par:tipsreconstruction} we devise two alternative approaches based on quantum tomography that allow  to reconstruct unknown time-independent master equation parameters. In section \ref{par:specific case} we apply these procedures to a benchmark model and compute the amount of measurements needed. In section \ref{par:Summary and Conclusions} we summarize and discuss our results.
\ref{TC} contains more details on the reconstruction of the cumulants of a Gaussian state  through symplectic tomography.

\section{From tomograms to cumulants}\label{par:tomogramscumulants procedure}

In \cite{Bellomo2009} we have introduced  a procedure that allows to reconstruct, via a limited number of measurements, the time-independent master equation coefficients governing the dynamical dissipative evolution of a quantum system. In the Markovian case, it has been shown that, using a Gaussian probe, the required number of measurements is at most ten.
 In facts, the dynamical evolution of a Gaussian state is completely determined by the evolution of its first and second order cumulants, which are measurable quantities.  The unknown master equation coefficients enter the dynamical equations of the cumulants, hence can  be retrieved by simple inversion, once the latter are measured. The cumulants can be obtained by using symplectic tomography. Indeed, given the Wigner function of a Gaussian state, it can be measured along lines in phase space (i.e. by performing its tomographic map). This allows to   relate the cumulants to  points on the tomogram. By choosing the lines in phase space corresponding, respectively, to  position and momentum probability distributions, one needs at most four points along each tomogram (i.e. line) to retrieve the first and second cumulant of the associated variable. The same procedure, applied on a line inclined by $\pi/4$ in phase space allows to retrieve the covariance of the two variables by measuring at most two points. Hence, given a Gaussian Shape Preserving (GSP) Markovian master equation, by using Gaussian probes one can retrieve at any time its evolved cumulants via a finite amount of measurements. In the following, we will refer to this  as the tomograms-cumulants (T-C) procedure. More details of the T-C procedure are provided in \ref{TC}. As the key ingredient of the T-C procedure is the preservation of Gaussianity, it is therefore a natural step to investigate how this procedure can be generalized and extended to more involved non-Markovian scenarios, still preserving Gaussianity. Indeed our aim is further supported by recent work \cite{Busch2008,Kiukas2009,Lahti2009}, in which it has been proved that it is in principle possible to make tomographic measurements of the probability densities associated to every quadrature in phase space (for example in quantum optics it could be realized by means of homodyne detection). As a final remark we note that other methods to measure the covariance matrix of Gaussian states have been discussed in \cite{Rehacek2009}. However in this case the amount of required measurements is much higher, the focus being  on the reduction of experimental errors.

\section{Non-Markovian master equation}\label{par:Description of the system}

We will focus on the class of master equations of the form \cite{Petruccione-Breuerlibro2002}:

\begin{eqnarray}\label{quasi lindblad form}
\fl \frac{d \hat{\rho}(t)}{d t}=L(t)[\hat{\rho}(t)]=-\frac{i}{\hslash}\left[\hat{H}, \hat{\rho}(t) \right]+\frac{1}{2\hslash}\sum_j
\left(\left[\hat{V}_j(t) \hat{\rho}(t),\hat{V}_j^\dag (t)
\right]+\left[\hat{V}_j (t), \hat{\rho}(t)\hat{V}_j^\dag (t)\right]\right),
\end{eqnarray}
where the generator $L(t)[\cdot]$  depends on time. In particular we shall investigate the following class of  time-dependent master equations 
\begin{eqnarray}\label{master equation in q e p}
\fl \frac{\mathrm{d} \hat{\rho}(t)}{\mathrm{d}
t}&=&\!-\frac{i}{\hslash}\left[\hat{H}_0,\hat{\rho}(t)
\right]-\frac{i (\lambda(t) +\delta)}{2
\hslash}\left[\hat{q},\hat{\rho} (t) \hat{p}+\hat{p}\hat{\rho}(t) \right]
 +\frac{i (\lambda(t) -\delta)}{2 \hslash}\left[\hat{p},\hat{\rho}(t)
\hat{q}+\hat{q}\hat{\rho}(t) \right]\nonumber\\ \fl &&-\frac{D_{pp}(t)}{
\hslash^{2}}\left[\hat{q},[\hat{q},\hat{\rho}] \right]
-\frac{D_{qq}(t)}{ \hslash^{2}}\left[\hat{p},[\hat{p},\hat{\rho}(t)]
\right]+\frac{D_{qp}(t)}{ \hslash^{2}}
\left(\left[\hat{q},[\hat{p},\hat{\rho}(t)]\right]+\left[\hat{p},[\hat{q},\hat{\rho}(t)]
\right]\right).
\end{eqnarray}
The master equation (\ref{master equation in q e p}) is  obtained from the general form (\ref{quasi lindblad form}) by choosing a system Hamiltonian of the form
\begin{equation}\label{Hamiltonian}
    \hat{H}=\hat{H}_0+\frac{\delta}{2}\left(\hat{q}\hat{p}+\hat{p}\hat{q}\right), \qquad \hat{H}_0= \frac{1}{2m}\hat{p}^2+\frac{m\omega^2}{2}\hat{q}^2,
\end{equation}
and  linear Lindblad operators $\hat{V}_j (t)$
\begin{equation}\label{linbblad operators}
    \hat{V}_j (t)=a_j (t) \hat{p} + b_j (t) \hat{q}, \qquad j=1,2.
\end{equation}
Using equations (\ref{Hamiltonian}) and (\ref{linbblad operators}) in (\ref{quasi lindblad form}), the time-dependent master equation coefficients in (\ref{master equation in q e p}) hence read
\begin{eqnarray}\label{D coefficients}
 && D_{qq}(t)=\frac{\hslash}{2}\sum_{j=1,2}|a_j(t)|^2, \qquad D_{pp}(t)=\frac{\hslash}{2}\sum_{j=1,2}|b_j(t)|^2, \nonumber \\ & & D_{qp}(t)=-\frac{\hslash}{2}\mathrm{Re}\sum_{j=1,2}a_j^*(t)b_j(t), \qquad \lambda(t)=-\mathrm{Im}\sum_{j=1,2}a_j^*(t)b_j(t),\end{eqnarray}
where the $D_{ij}(t)$'s and  $\lambda(t)$ are, respectively, the so-called diffusion and friction coefficients.
The master equation in  (\ref{master equation in q e p}) represents a natural generalization of the time-independent master equation introduced in \cite{Sandulescu1987} which describes a
GSP evolution of a quantum state. In the time-dependent case a wide range of models obeys a GSP master equation of the form (\ref{master equation in q e p})
\cite{HM1994,HR1985,HY1996,Strunz2004,Yu2004,Bassi2009,KG1997,RP1997}.
As a side remark we note here that the literature about non-Markovian master equations may lead to some ambiguity. Indeed some authors
classify as non-Markovian only those master equations  whose generator contains a convolution integral.  It has recently been proved in \cite{Kossakowski2010} that these generators can be mapped into convolutionless ones, following a so-called local approach. Non-Markovianity becomes then characterized by the dependence of the convolutionless generator on $t-t_0$ where $t_0$ is the initial time. According to this
approach a time-dependent convolutionless generator as equation (\ref{master equation in q e p}) could be considered
Markovian. However, following a consistent part of literature, e.g. \cite{HPZ1992, HR1985, HY1996, Strunz2004,Yu2004,HPZ1993,Maniscalco2004}, we  term non-Markovian convolutionless time-dependent generators as the one in (\ref{master equation in q e p}).

\subsection{Evolution of the cumulants} \label{appA}

We begin by writing the dynamical equations, obtained from (\ref{master equation in q e p}), for the cumulants of a Gaussian state. They can be expressed in compact matrix form as
\begin{eqnarray}
& & \frac{d}{dt}S(t)=\left(M-\lambda(t) I_2\right)S(t), \label{eq1} \\
& & \frac{d}{dt}X(t)=\left(R-2\lambda(t) I_3\right)X(t)+D(t), \label{eq2}
\end{eqnarray}
where $I_{2(3)}$ is  the 2(3)-dimensional identity matrix. The vectors  $S(t)$ and $X(t)$ correspond, respectively, to the first and second order cumulants
\begin{equation}
 S(t)=\frac{1}{\sqrt{\hslash}}\left(\begin{array}{c}
 \sqrt{m\omega}\langle\hat q\rangle_t \\
 \frac{\langle\hat p\rangle_t}{\sqrt{m\omega}}
\end{array}\right),\qquad  X(t)= \frac{1}{\hslash}
\left(\begin{array}{c}
m\omega\Delta q_t^2 \\
 \Delta p^2_t/(m\omega)\\
(\sigma_{q,p})_t
\end{array}\right)
,\label{momenta}\end{equation}
the matrices $M$ and $R$ contain the Hamiltonian parameters
\begin{equation} M=\left(\begin{array}{cc}
\delta & \omega \\
-\omega & -\delta
\end{array}\right)
, \qquad R=\left(
\begin{array}{ccc}
2\delta & 0 & 2\omega \\
0 & -2\delta & -2\omega \\
-\omega & \omega & 0
\end{array}\right)\label{MR}\end{equation}
and, finally,  $D(t)$ is the diffusion vector
\begin{equation}
D(t)= \frac{2}{\hslash}\left(\begin{array}{c}
m\omega D_{qq} \\
\frac{D_{pp}}{m\omega} \\
 D_{qp} \\
\end{array}\right). \label{D}\end{equation}
As in the time-independent case, the first cumulant dynamical evolution (\ref{eq1}) depends only on  $\lambda(t)$ also entering, together with  the diffusion coefficients, the second order cumulants  equation (\ref{eq2}). Hence, the friction coefficient can  be retrieved by inverting (\ref{eq1}). By carrying out a formal integration one obtains the following expression
\begin{equation} \label{lambda}
\int_0^t dt'\lambda(t')=\ln\left(\frac{\tilde S_j(0)}{\tilde S_j(t)}\right),
\end{equation}
where the suffix $j=1,2$ labels the two components of the vector \begin{equation}
\tilde S(t)=e^{-tM}S(t).\label{S}\end{equation}
Analogously,  (\ref{eq2}) can be rewritten as \begin{equation}
\frac{d}{dt}\tilde X(t)=\tilde D(t),
\label{diff2}\end{equation}
where
\begin{equation}
\tilde X(t)=e^{2\int_0^t dt'\lambda(t')}e^{-t R}X(t), \qquad
\tilde D(t)=e^{2\int_0^t dt'\lambda(t')}e^{-t R}D(t).
\label{XD}
\end{equation}
It can be shown \cite{Sandulescu1987} that the  transformations in (\ref{S}) and (\ref{XD}) are always invertible, provided one sets the quantity $\eta\equiv\sqrt{\delta^2-\omega^2}$ to  $i\Omega$ whenever $\eta^2<0$.
The formal solution of (\ref{diff2}) is given by
\begin{equation}
\tilde X(t)=\tilde X(0)+\int_0^t dt'\tilde D(t').
\end{equation}
Inverting the transformation in (\ref{XD}), one can write

\begin{eqnarray} \label{D}\int_0^t dt'e^{-2\int_{t'}^t dt''\lambda(t'')} e^{(t-t') R}D(t')= X(t)-e^{tR}e^{-2\int_0^{t} dt''\lambda(t'')}X(0).
\end{eqnarray}Both in (\ref{lambda}) and (\ref{D}), measurable quantities appear on the right hand side, whereas the (unknown) coefficients are on the left.

\section{Reconstruction of time independent parameters}\label{par:tipsreconstruction}

Hereafter we will assume that the master equation coefficients (MECs) $\lambda(t), D_{qq}(t), D_{pp}(t)$ and $D_{qp}(t)$  have a known functional form. This implies that the non-Markovian master equation with certain expressions for the MECs has been previously derived within some approximation scheme (e.g. by means of a microscopic derivation and subsequent dynamical assumptions). The time-dependent MECs  are thus function of a set of  time-independent  parameters (TIPs) whose value is {\it a priori} unknown. In this section we propose two alternative procedures aiming at reconstructing the TIPs by means of symplectic tomography.

\subsection{Integral approach}\label{sec: integral approach}

Here we introduce an approach based on formal integration of the dynamical equations (\ref{lambda}) and (\ref{D}).
The right hand side of both equations involves experimental inputs ($S(t)$ and $X(t)$) and known Hamiltonian parameters ($R$). The left hand sides, once the MECs are known, can be regarded as functions of the TIPs. Hence the set of TIPs can be  in principle obtained by inverting these relations.
Unfortunately, in general, an analytical inversion of (\ref{lambda}) and (\ref{D}) may represent a highly involved task. In facts, even if we assume a known MECs time-dependence, we could be unable to either compute analytically the integrals on the right hand side of (\ref{lambda}) and (\ref{D}), or to invert the equations or even both. All these problems can be anyway overcome by resorting to numerical computation. To provide an example of how to apply  this procedure, in the next section we will apply it to a specific model.

\subsection{Differential approach}\label{sec: differential approach}

The tomographic T-C procedure recalled in section \ref{par:tomogramscumulants procedure} allows us to measure not only the cumulants of a given Gaussian state but also their first time derivatives. Indeed, we can estimate the derivative through the  incremental ratio by measuring each cumulant at two different times $t$ and  $t+\delta t$. For instance
\begin{equation}\label{rappIncr}
\frac{d}{dt}\Delta q_t^2\sim\frac{\Delta q_{t+\delta t}^2-\Delta q_t^2}{\delta t},
\end{equation}
where the amount of time $\delta t$ is defined as the smallest time interval which allows to experimentally distinguish two different values of the given cumulant. Once we substitute derivatives with their approximations, the two sets of equations (\ref{eq1}) and (\ref{eq2}) are not differential anymore.
Being the cumulants and their approximate derivatives at given times experimental inputs, we insert in (\ref{eq1}) and (\ref{eq2}) the time-dependent coefficients $\lambda(t)$ and $D(t)$ (and in case $m(t)$, $\omega(t)$, $\delta(t)$), which involve the unknown TIPs and in this way the two sets of equations (\ref{eq1}) and (\ref{eq2}) reduce to  algebraic equations which can always be solved, at least numerically.
Due to this simplification, this approach is also suitable when dealing with more complicated generators than the one considered here in equation ({\ref{master equation in q e p}), e.g. when the Lamb shift contribution is explicitly taken into account such that the Hamiltonian parameters become time dependent: $m(t)$, $\omega(t)$, $\delta(t)$.} We note that  the two sets (\ref{eq1}) and  (\ref{eq2}) consist of five equations which must be fulfilled at any chosen time. Therefore, considering them at different times, we can derive a system made up of an arbitrary number of equations. The number of equations must then be chosen as the minimum amount of  equations needed to  uniquely determine the TIPs, which is clearly model dependent. In the following section, we show how to apply this procedure to a benchmark model.

\section{A benchmark model}\label{par:specific case}

In the following we will refer to a specific model of a quantum Brownian particle discussed in \cite{Maniscalco2004}. We will show how to apply in this specific case the two general procedures presented in the previous section. The model  consists of an Ohmic reservoir made of harmonic oscillators, linearly coupled to a single harmonic oscillator of frequency $\omega$ (our system particle) through the coupling constant $\alpha$, with a Lorentz-Drude cut-off \cite{Petruccione-Breuerlibro2002} $\omega_c$ and at temperature $T$. Starting from a superoperatorial version of the Hu-Paz-Zhang master equation \cite{HPZ1992}, a secular master equation of the form (\ref{master equation in q e p})  is obtained in the weak-coupling limit (up to the second order in $\alpha$), with the following coefficients:
\begin{eqnarray}\label{lambda t}
    \quad\delta & =&0, \qquad D_{qp}=0, \qquad \frac{m \omega D_{qq}}{\hslash}=\frac{D_{pp}}{\hslash m \omega}=\frac{\Delta(t)}{2},\nonumber \\
     \lambda(t)&=&\frac{\alpha^2 \omega_c^2 \omega}{ \omega_c^2 + \omega^2} \left\{
    1-e^{-\omega_c t} \left[\cos (\omega t)+ \frac{\omega_c}{\omega} \sin (\omega t)\right]\right\},\nonumber \\
    \Delta(t)&=&\frac{ 2 \alpha^2   \omega_c^2  }{\omega_c^2 +\omega^2} \frac{k T}{\hslash}\left\{
    1-e^{-\omega_c t} \left[\cos (\omega t)- \frac{\omega}{\omega_c}\sin (\omega t)\right]\right\},
\end{eqnarray}
the last having this form  for high temperatures $T$. This master equation is
of Lindblad-type when the coefficients $\Delta(t)\pm \lambda(t)$ are positive
at all times. The Lindblad-non Lindblad border as a function of the temperature $T$ and the frequency cutoff $\omega_c$  has been analyzed in \cite{Maniscalco2004}. As Gaussianity is preserved, by choosing a Brownian particle  initially in a Gaussian state, the T-C procedure can be employed at any time. The coefficients $ \lambda(t)$ and  $\Delta(t)$ reach stationary values for $t\gg 1/\omega_c$
\begin{equation}\label{stationary values}
\lambda(t)\rightarrow \frac{\alpha^2 \omega_c^2 \omega}{ \omega_c^2 + \omega^2},\qquad \Delta(t)\rightarrow \frac{ 2 \alpha^2   \omega_c^2  }{\omega_c^2 +\omega^2} \frac{k T}{\hslash}.
\end{equation}
In this specific model, the unknown TIPs are the coupling constant $\alpha$, the temperature $T$ and the frequency cut-off $\omega_c$. Usually, when studying quantum Brownian motion, one assumes $\omega_c/\omega \gg 1$, corresponding to a  Markovian reservoir, with $\omega_c \rightarrow \infty$. In this limit, the thermalization time \cite{Maniscalco2004} is inversely proportional to the coupling strength, while for an out-of-resonance engineered reservoir  with  $\omega_c/\omega \ll 1$ (i.e. highly non Markovian), the thermalization process is slowed down.

\subsection{Example: integral approach}

\begin{figure}
 \centering
\subfigure[]
{\label{1markov}\includegraphics[width=0.45\columnwidth]{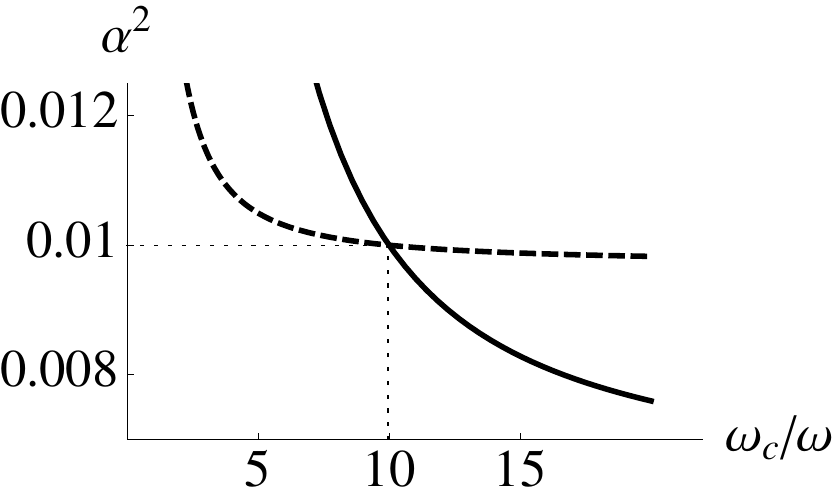}
}
 \subfigure[]
   {\label{1nonmarkov}\includegraphics[width=0.45\columnwidth]{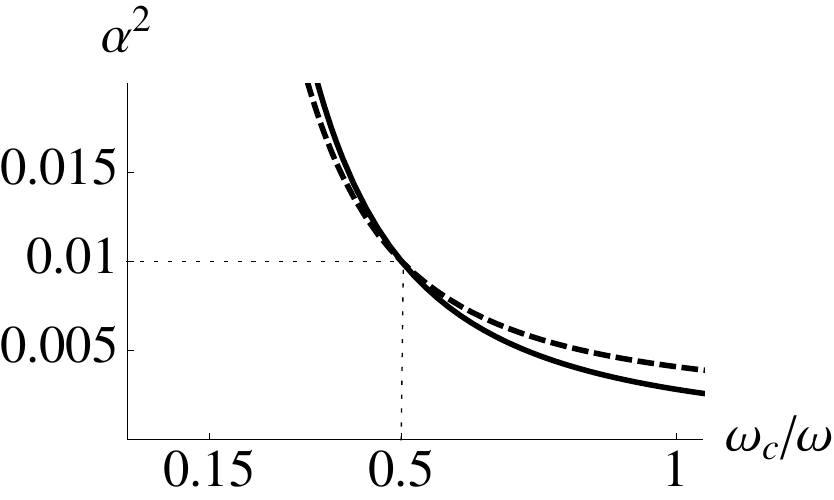}}{\label{fig1b}}
\caption{We show how to indirectly measure the time independent parameters $\omega_c$ and $\alpha^2$, in two different regimes respectively close to
 the Markovian (figure \ref{1markov}) and non-Markovian dynamics (figure \ref{1nonmarkov}). Each line refers to an experimental measure of  $ \ln\left(\tilde S_j(0)/\tilde S_j(t)\right)$ at a specific time $\omega t$. In both regimes, the time independent parameter values are found at the intersection of the two lines. In the example close to the Markovian regime (figure \ref{1markov}) if we measure $3.03\cdot10^{-3}$ at $\omega t_1=0.5$ (solid line) and $9.70\cdot10^{-2}$ at $\omega t_2=10$ (dashed line), we retrieve $\alpha^2=0.01$ and $\omega_c/ \omega=10$. Analougsly, for the example close to the non Markovian regime (figure \ref{1nonmarkov}) if we measure  $4.55\cdot10^{-5}$ at $\omega t_1=0.5$ (solid line) and $1.84\cdot10^{-2}$ at $\omega t_2=10$ (dashed line), we retrieve $\alpha^2=0.01$ and $\omega_c/ \omega=0.5$.}
\label{fig:integral}
 \end{figure}

Here we apply the integral procedure (section \ref{sec: integral approach}) to the benchmark model. In this case the left hand side of (\ref{D}) is not simply analytically computable. Thus we must use (\ref{lambda}) to reconstruct all the TIPs it involves, and then numerically integrate the left hand side of (\ref{D}). \\
The left hand side of (\ref{lambda}) is given by

\begin{eqnarray} \label{lambda int} \fl \int_0^t dt'\lambda(t')=\frac{\alpha^2  \omega_c^2 \omega^2}{(\omega_c^2+\omega^2)^2}\left\{\omega t \frac{ \omega_c^2 +\omega^2}{\omega^2}-2\frac{\omega_c}{\omega}\right.
\left.+e^{-\omega_c t} \left[2 \frac{\omega_c}{\omega} \cos(\omega t)+\frac{ \omega_c^2 -\omega^2}{\omega^2}
\sin(\omega t)\right]\right\}. \nonumber\\
\end{eqnarray}
By using (\ref{lambda}) and (\ref{lambda int}) we obtain the following trascendental equation for the coupling strength $\alpha$
\begin{eqnarray}
\alpha^2  &=&  \ln\left(\frac{\tilde S_j(0)}{\tilde S_j(t)}\right)\frac{(\omega_c^2+\omega^2)^2   }{\omega_c^2 \omega^2 }\left\{\omega t \frac{ \omega_c^2 +\omega^2}{\omega^2}-2\frac{\omega_c}{\omega}\right. \nonumber \\
&&\quad \left.+e^{-\omega_c t} \left[2 \frac{\omega_c}{\omega} \cos(\omega t)+\frac{ \omega_c^2 -\omega^2}{\omega^2}
\sin(\omega t)\right]\right\}^{-1},
\label{alpha}
\end{eqnarray}
where the ratio $\tilde S_j(0)/\tilde S_j(t)$ is the experimentally measurable quantity. Hence, by performing two distinct measurements of this ratio we can evaluate (\ref{alpha}) at two different times. We thus obtain  a system of two numerically solvable equations,  which allows us to retrieve  the time-independent parameters  $\alpha$  and $\omega_c$.  To provide a concrete evidence of the validity of this procedure, we show two numerical examples in figure \ref{fig:integral}.  Indeed we retrieve the TIP $\alpha^2=0.01$ in two different dynamical regimes, respectively close to the Markovian (figure \ref{1markov}) and non-Markovian (figure \ref{1nonmarkov}) limit.
The last missing parameter is the temperature $T$ entering the coefficient $\Delta(t)$. By using (\ref{D}), we obtain   the following equation:
\begin{eqnarray}
\label{temp}
\fl \frac{k T}{\hslash \omega} &=&  \left[X_j(t)-e^{-2\int_0^{t} dt''\lambda(t'')}\left(e^{tR}X(0)\right)_j\right]  \frac{\omega_c^2+\omega^2}{2\alpha^2  \omega^3}\Bigg\{\int_0^t dt'e^{-2\int_{t'}^t dt''\lambda(t'')}\sum_{l=1}^2\left(e^{(t-t') R}\right)_{j,l} \nonumber \\
\fl & & \times \left\{
    1-e^{-\omega_c t'} \left[\cos (\omega t')- \frac{\omega}{\omega_c}\sin (\omega t')\right]\right\}\Bigg\}^{-1},
\end{eqnarray}
where $j=1,2,3$ denotes the vector components, and $\left(e^{(t-t') R}\right)_{j,l}$ are the matrix elements of the matrix $e^{(t-t') R}$.
The explicit expression of the integral appearing on the first line of (\ref{temp}) is provided in (\ref{lambda int}). In  general, the remaining integrals are not analytically computable. However, since all the parameters involved have been previously reconstructed, these integrals can be computed numerically.

\subparagraph{Number of tomographic measurements}

Let us now explicitly compute the number of tomograms  needed to apply the integral approach  to the benchmark model. To reconstruct $\alpha$ and $\omega_c$   each of the quantities $\tilde S_{1,2}(t)=\left(e^{-tM}S\right)_{1,2}(t)$ must be measured once but not at the same time, as shown by  (\ref{alpha}) and figure \ref{fig:integral}. Each $\tilde S_{1,2}(t)$ is a function of the first cumulants of both position and momentum.  According to the T-C procedure (see \ref{TC}), the reconstruction of a first cumulant involves at most four  tomographic points. Thus $\alpha$ and $\omega_c$ can be obtained via, in the worst case, sixteen measures.  Furthermore, being $e^{-tM}$  an orthogonal transformation, $e^{-tM}S_j(t)$ is by itself a first cumulant along a time-dependent direction in  phase-space. Hence, if  time-dependent tomographic measurements (i.e. measurements in a frame rotating as $e^{-tM}$) are allowed, the number of required tomograms  decreases to eight, as we would only need a single first cumulant  ($\tilde S_1(t)$ or $\tilde S_2(t)$).\\
To measure  $T$ we should evaluate one of the second cumulants at a given time. Following the T-C procedure this amounts to two tomographic points. However, the required second cumulant has been already obtained when reconstructing the corresponding first cumulant, hence the temperature can be retrieved without further effort. This argument also holds for time-dependent measurements. In facts, the reconstruction formula (\ref{temp}) has been derived from (\ref{D}), which  can be recast in terms of the variances in the rotating frame. The temperature  can be then obtained using the variance along the same time-dependent direction of  the measured first cumulant.

In conclusion, in order to implement the integral approach in the benchmark model, according to whether we can perform time-dependent measurements or not, we need eight or at most sixteen tomographic points.

\subsection{Example: differential approach}

\begin{figure}
 \centering
\subfigure[]
{\label{2markov}\includegraphics[width=0.45\columnwidth]{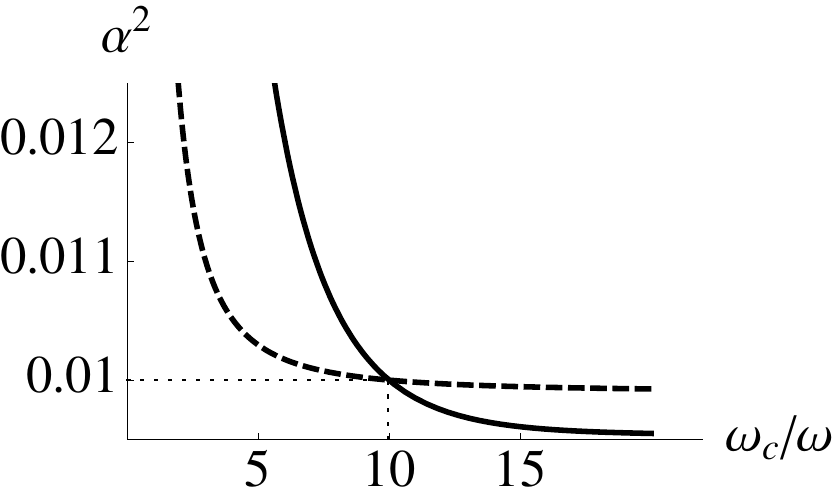}
}
\subfigure[]
{\label{2nonmarkov}\includegraphics[width=0.45\columnwidth]{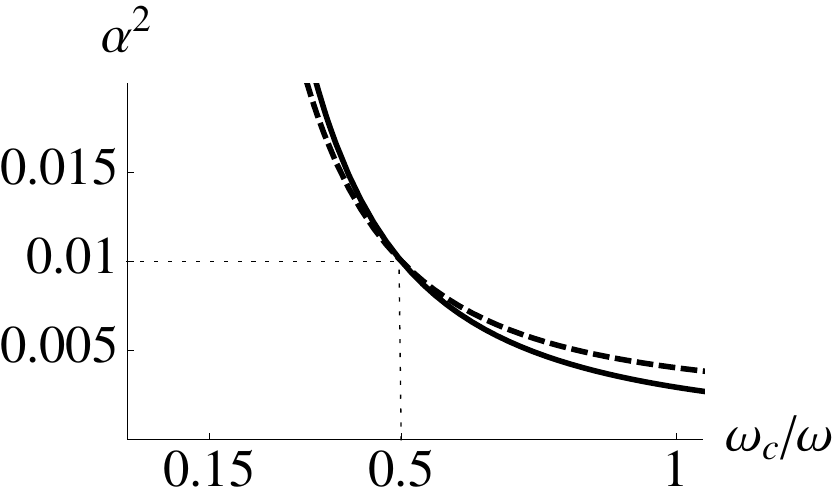}}
\caption{  As for the integral approach we obtain an indirect measure of the time-independent parameters $\omega_c$ and $\alpha^2$, both in the almost Markovian (figure \ref{2markov}) and in the almost non-Markovian regime (figure \ref{2nonmarkov}). Each line refers to an experimental measure of  $\frac{1}{\omega \langle\hat{q}\rangle_t }\left(\frac{\langle\hat{p}\rangle_t}{m}-\frac{\langle\hat{q}\rangle_t- \langle\hat{q}\rangle_{t+\delta_t}}{\delta t}\right)$ at a specific time $\omega t$. In the almost Markovian example shown in figure \ref{2markov}, if we obtain $9.52 \cdot 10^{-3}$ at $\omega t_1=0.5$ (solid line) and $9.90\cdot10^{-3}$ at $\omega  t_2=10$ (dashed line), we retrieve $\alpha^2=0.01$ and $\omega_c/ \omega=10$. Analogously, for the almost non-Markovian case in  figure \ref{2nonmarkov}, if we measure $2.59\cdot10^{-4}$ at $\omega t_1=0.5$ (solid line) and $2.01\cdot10^{-3}$ at $\omega t_2=10$ (dashed line), we retrieve $\alpha^2=0.01$ and $\omega_c/ \omega=0.5$. }
\label{fig:differential}
\end{figure}

We now skip to the differential procedure (section \ref{sec: differential approach}).
In (\ref{lambda t}) the dependence on $\alpha^2$ and $\omega_c$ is factorized, hence using
(\ref{eq1}) one gets
 \begin{eqnarray}
 \fl \alpha^2  \sim  \frac{1}{\langle\hat{q}\rangle_t }\left(\frac{\langle\hat{p}\rangle_t}{m}-\frac{\langle\hat{q}\rangle_t- \langle\hat{q}\rangle_{t+\delta_t}}{\delta t}\right)\frac{ \omega_c^2 + \omega^2 }{\omega_c^2}  \frac{1}{\omega} \left\{
    1-e^{-\omega_c t} \left[\cos (\omega t)+ \frac{\omega_c}{\omega} \sin (\omega t)\right]\right\}^{-1}. \label{A}
 \end{eqnarray}
Since $\alpha^2$ and $\omega_c$ are time-independent, they can be determined by solving (\ref{A}) for two different times $t_1$ and $t_2$, and looking at the intersection of the two different solutions. This procedure requires to measure the cumulants $\langle\hat{q}\rangle_t$, $\langle\hat{q}\rangle_{t+\delta_t}$ and $\langle\hat{p}\rangle_{t}$ at $t=t_1, t_2$  and  to solve (\ref{A}) numerically, as shown in figure \ref{fig:differential}. In other  words, the first two TIPs, $\alpha^2$ and $\omega_c$, can be determined by measuring six quantities. As for the integral procedure, we retrieve the TIP $\alpha^2=0.01$ in two cases, corresponding to the extreme dynamical regimes,  Markovian  $\omega_c/\omega\gg 1$ in figure \ref{2markov}, and the highly non-Markovian, $\omega_c/\omega\ll 1$, in  figure \ref{2nonmarkov}.
Again, we are left with determining the temperature $T$ appearing in  (\ref{lambda t}). To this end, we consider one of  the equations of system (\ref{D}), e.g. that  for $(2/\hbar)m\omega D_{qq}(t)$, which according to (\ref{rappIncr}) we reformulate as:
\begin{eqnarray}\label{rappIncrForT}
&&\frac{k T}{\hslash  \omega}
\sim \frac{1}{\hslash}\left[m \left(\frac{\Delta q_{t+\delta_t}^2 -\Delta q_t^2}{\delta t} + 2 \lambda (t) \Delta q_t^2 \right) -2 \sigma(q,p)_t \right]\nonumber \\
&&  \qquad  \qquad  \times \frac{  \omega_c^2 +\omega^2 }{  2 \alpha^2   \omega_c^2  } \left\{
    1-e^{-\omega_c t} \left[\cos (\omega t)- \frac{\omega}{\omega_c}\sin (\omega t)\right]\right\}^{-1}.
\end{eqnarray}
Equation (\ref{rappIncrForT}) allows to retrieve $T$ once the cumulants $\Delta q_t^2$,  $\Delta q_{t+\delta_t}^2$ and $\sigma(q,p)_t$ are measured and both $\alpha^2$ and $\omega_c$ are known from the previous steps.
As an example, at time $\omega t=1$  the same value of temperature, $kT=10\hslash\omega$  may correspond to  different measured values of  the quantity
 $1/\hslash\left[m \left((\Delta q_{t+\delta_t}^2 -\Delta q_t^2)/\delta t + 2 \lambda (t) \Delta q_t^2 \right) -2 \sigma(q,p)_t \right]$, according to different dynamical regimes. For instance, by setting $\omega_c/\omega=10$ close to the Markovian regime one would measure $0.198$ as output whereas, setting  $\omega_c/\omega=0.5$ close to the non-Markovian case, the output corresponding to the same $T$ is  $0.067$.

\subparagraph{Number of tomographic measurements}

We now count the number of tomographic measurements to apply the differential approach to this example. The reconstruction of both $\alpha$ and $\omega_c$ is  based on (\ref{A}), which must then be evaluated at two different times $t_1$ and $t_2$, see figure \ref{fig:differential}. Each  evaluation of (\ref{A}) requires two measurements of  the average position, at times $t_i$ and $t_i+\delta t$, and one of the average momentum at time $t_i$, where $i=1,2$. This implies reconstructing  six first cumulants. As each first cumulant requires four tomographic points (see \ref{TC}), the total amount of needed tomographic  points amounts to twenty-four.\\
The reconstruction of  temperature $T$ is based on equation (\ref{rappIncrForT}). We need  the variance of the position at time $t$ and at time $t+\delta t$, and the covariance of $\hat{q}$ and $\hat{p}$ at time $t$. However, according to the T-C procedure (\ref{TC}),  the variance is required to obtain the position average. This implies that the second cumulant has been already measured during  the previous reconstruction, and there is no need to measure it again. Retrieving  the  covariance requires two more tomographic points. \\
In conclusion, in order to implement the differential approach in the benchmark model, we need twenty-six tomograms.
\subparagraph{Comparison between the two approaches}
 Let us now briefly compare the two procedures described in this section.
On one hand the differential approach requires more
experimental measurements compared to the integral one while, on the other,
the latter procedure is more involved from a computational point of
view. Indeed, it may happen that to compute the first members of
Eqs.~(11)-(16) some numerical or analytical approximations are needed, thus
reducing the accuracy of  the reconstruction. In this case, the differential approach should be preferred
as it is very simple from the point of view of analytical computation. Clearly, if the
computation of the integral functions in Eqs.~(11)-(16) does not present
remarkable difficulties, the integral procedure proves better as, requiring less measurements, it involves a lower number of interactions with the physical system. For example, in our benchmark model, the differential approach requires
twenty-six measurements, while the integral approach requires sixteen
time-independent measurements or only eight time-dependent measurements.\\
One could summarize by saying that the integral procedure is more advantageous in terms of number of measurements, but requires the ability of solving potentially involved analytical expressions. The differential approach, instead, is more advantageous from the point of view of versatility, as it allows to deal in a straightforward way with complex generators, at the expenses of a higher number of measurements.
In conclusion, the choice between the two strategies introduced
in this paper strictly depends on the specific model under
investigation.

\section{Conclusions}\label{par:Summary and Conclusions}

In this paper we have proposed an experimentally feasible procedure to reconstruct unknown time-independent master equation parameters in a non-Markovian scenario. To this end we have adopted an approach based on symplectic tomography.
While previously the case of Markovian dynamics has been investigated \cite{Bellomo2009}, here the procedure is generalized and extended  to the more involved convolutionless non-Markovian case.
In particular, our analysis is focused on  the class of Gaussian Shape Preserving master equations. We have addressed the situation in which the time-dependent  master equation coefficients are  analytic functions of  some unknown time-independent parameters. The key point of our approach lies in using Gaussian states as probes, as  information on the dissipative dynamics can be inferred via a limited number of tomograms.  We have proposed two alternative procedures, integral and differential, to reconstruct the unknown quantities. In order to provide an explicit example of how these different approaches work, we have applied them to a benchmark model made up of a harmonic oscillator coupled to a bosonic bath,  whose unknown parameters are the coupling constant, the temperature and the bath frequency cut-off. In this case the number of needed tomograms ranges between sixteen (at most) in the integral approach and  twenty-six  in the differential approach. \\
Besides measuring unknown parameters, our procedure proves useful also in case those are already known. Indeed it could be employed as a preliminary consistency test for the adopted master equation. In facts, the reconstruction procedure assumes that the time-dependent master equation coefficients are previously known  functions of a set of time-independent quantities.  This is for example the case of a microscopical derivation (and related approximations) of the master equation. In this perspective, the agreement between the measured and theoretically expected time-independent parameters  provides a necessary validity condition for  the adopted approximation scheme.  Along the same line of thought one could also envisage an extension of  this approach to  the reconstruction of the whole set of time-dependent master equation coefficients. This could provide a sound, reliable and complete experimental check of the goodness of the approximation scheme underlying a master equation. A detailed investigation of this wider scenario will be subject of a distinct
study \cite{Bellomo2010}. Our proposal opens up several interesting questions which are going to be the subject of further future investigation.  In facts  how our approach can be recast within an estimation theory perspective represents a relevant open scenario. A similar analysis has indeed been performed in \cite{Alex2007} for a single-parameter Markovian master equation. Another relevant point  to investigate is whether the proposed protocol can be enhanced by employing entangled Gaussian states as a probe. Finally,
whether or not the proposed procedure can be generalized and employed in presence of memory kernels is a challenging question. Indeed, reconstructing the unknown parameters of Gaussian noisy evolutions with memory represents both a highly involved and interesting task.

\ack

We warmly thank Paolo Facchi and Marco Lucamarini for useful discussions. A. D. P. acknowledges  financial
support from the European Union through the Integrated Project
EuroSQIP.

\appendix
\section{The T-C procedure}\label{TC}

Tomographic maps \cite{Lvovsky2009,Asorey2007} allow to reconstruct the state or some other properties of a physical system, both in a classical
and in a quantum regime. In general, tomography-based  techniques stem from a probabilistic perspective.
Indeed, given a quantum state $\hat{\rho}(t)$ its Wigner function \cite{Wigner1932, Moyal949} provides a
generalization on phase space of a classical probability distribution and   is  defined as
\begin{equation}\label{wigner function definition}
    W(q,p,t)=\frac{1}{\pi \hslash} \int_{-\infty}^{+\infty}\mathrm{d}y \exp \left(
    \frac{i 2 p y}{\hslash}\right)\hat{\rho} (q-y,q+y,t). \\
\end{equation}
The previous equation can be read as a map between real phase-space  functions  and density matrices. In particular, whenever the  dynamics of a quantum system initially in a Gaussian state obeys the master equation
 (\ref{master equation in q e p})  
 the associated Wigner function is a Gaussian function itself and reads
\begin{eqnarray}\label{wigner function gaussian}
\fl W(q,p,t)  =&&  \frac{1}{2 \pi \sqrt{\Delta q_t^2\Delta p_t^2-\sigma(q,p)_t^2}}
   \exp \Bigg[-  \frac{\Delta q_t^2
  (p-\langle \hat{p}\rangle_t)^2+\Delta p_t^2(q-\langle \hat{q}\rangle_t)^2}{2[\Delta q_t^2\Delta p_t^2-\sigma(q,p)_t^2]}\nonumber\\&& \qquad \qquad \qquad \qquad-\frac{2\sigma(q,p)_t (q-\langle \hat{q}\rangle_t)(p-\langle \hat{p}\rangle_t)}{2[\Delta q_t^2\Delta p_t^2-\sigma(q,p)_t^2]} \Bigg]
  \,.
\end{eqnarray}
Given the Wigner distribution of a  quantum system, the Radon transform \cite{Radon1917} represents the key ingredient to perform a tomographic analysis. This invertible integral transformation allows to retrieve the marginal probability densities of the system, i.e. the probability density along straight lines in   phase space:
\begin{equation}\label{line in q p plane}
    X-\mu q - \nu p = 0.
\end{equation}
The formal expression of the Radon transform, for a generic quantum state, is then given by
\begin{eqnarray}\label{radon transform}
  \varpi (X,\mu,\nu)=\langle \delta\left(X-\mu q - \nu p\right)
    \rangle=
    \int_{\mathbb{R}^2} W(q,p,t)  \delta\left(X-\mu q - \nu p\right) \mathrm{d}q
    \mathrm{d}p.\nonumber\\
\end{eqnarray}
In case of the Gaussian function (\ref{wigner function gaussian}) it becomes:
\begin{eqnarray}\label{radon transform gaussian}
\fl  \varpi (X,\mu,\nu)=
    \frac{1 }{\sqrt{2\pi}\sqrt{\Delta q_t^2\mu^2+\Delta p_t^2 \nu^2 + 2\sigma(q,p)_t \mu \nu}}\exp \left[-\frac{\left(X-\mu \langle \hat{q}\rangle_t - \nu \langle \hat{p}\rangle_t \right)^2}{2[\Delta q_t^2\mu^2+\Delta p_t^2 \nu^2 + 2\sigma(q,p)_t \mu \nu]}
    \right].\nonumber\\
\end{eqnarray}
The second cumulants always obey the constrain $
\Delta q_t^2\mu^2+\Delta p_t^2 \nu^2 + 2\sigma(q,p)_t \mu \nu
>0$ as a consequence of the Robertson-Schr\"odinger relation \cite{Robertson1934}. This matrix inequality is
a generalization of the Heisenberg principle. The advantage of using Gaussian
probes to investigate a dissipative dynamics arises also within a statistical perspective.
 One could in fact wonder whether, due to experimental errors,  a violation of the
uncertainty principle might be observed. This may
happen if measurements are performed on states almost saturating the
Robertson-Schr\"odinger inequality, i.e. on the minimum uncertainty states which are pure.
However, our measurements are performed
on states undergoing a dissipative evolution i.e. on states typically far from being
pure hence from saturating the inequality.  Furthermore
 any additional noise of statistical origin will
have the effect of moving the reconstructed state further away from the boundary,
as noted in \cite{Rehacek2009}.

Let us  now consider the tomograms corresponding to the position and
momentum probability distribution functions ($X=q$ and $X=p$):
\begin{eqnarray}
\label{pdfq}
\varpi(X,1,0) & = & \frac{1}{\Delta q_t\sqrt{2\pi}}\exp \left[-\frac{\left(X-\langle \hat{q}\rangle_t\right)^2}{2\Delta q_t^2}\right],  \\
\label{pdfp}
\varpi(X,0,1) & = & \frac{1}{\Delta p_t\sqrt{2\pi}}\exp \left[-\frac{\left(X-\langle \hat{p}\rangle_t\right)^2}{2\Delta p_t^2}\right].
\end{eqnarray}
The lines individuated by the choices $(\mu,\nu)=(1,0)$ and $(\mu,\nu)=(0,1)$ correspond to
tomograms depending on the time average and variance respectively of
position and momentum. In order to determine the latter quantities
we must  invert (\ref{pdfq}) and (\ref{pdfp}) for different
values of $X$, i.e. for a given number of points to measure
along a tomogram. By  considering first   the direction $\mu=1$,
$\nu=0$, and by inverting (\ref{pdfq}),
we obtain:
\begin{equation}
\left(X-\langle \hat{q}\rangle_t\right)^2=2\Delta q_t^2\ln\frac{1}{\varpi(X,1,0)\Delta q_t\sqrt{2\pi}}\,.
\end{equation}
If we know the sign of $\langle \hat{q}\rangle_t$ then we
need only the value of the tomogram $\varpi(0,1,0)$ to get
$\langle \hat{q}\rangle_t$, otherwise we need another point. In this
way we get $\langle \hat{q}\rangle_t$ as a function of $\Delta
q_t$:
\begin{equation}\label{media di q from tomigram}
    \langle \hat{q}\rangle_t=\pm\Delta q_t\sqrt{2\ln\frac{1}{\varpi(0,1,0)\Delta q_t\sqrt{2\pi}}}\,.
\end{equation}
Using (\ref{media di q from tomigram}), (\ref{pdfq})
becomes an equation for $\Delta q_t$ only, and it can be rewritten
as
\begin{eqnarray} \label{eqtrascend} 2\Delta
q_t^2 \ln\frac{1}{\varpi(X,1,0)\Delta q_t\sqrt{2\pi}} =\left(X\mp\Delta q_t\sqrt{2\ln\frac{1}{\varpi(0,1,0)\Delta
q_t\sqrt{2\pi}}}\right)^2. \nonumber \\
\end{eqnarray}
This equation is trascendental, therefore we can only solve it
numerically. For each $X$ and
corresponding $\varpi(X,1,0)$ there may be two values of $\Delta
q_t$ satisfying the previous equation. In order to identify one of
the two solutions, it is enough to consider two points,
$\left\{(X_1,\varpi(X_1,1,0))\right\}$ and
$\left\{(X_2,\varpi(X_2,1,0))\right\}$, and to choose the common
solution for the variance.
Hence, whether we know or not the sign of the average $\langle
q\rangle_t$, we need three or four points to determine
$\langle \hat{q}\rangle_t$ and $\Delta q_t$ in (\ref{pdfq}).
Analogously, we need other three or four points for
$\langle \hat{p}\rangle_t$ and $\Delta p_t$ in (\ref{pdfp}).
We now compute the covariance
$\sigma(q,p)_t$. To this end we consider the tomogram:
\begin{eqnarray}
\fl \varpi\left(X,\frac{1}{\sqrt{2}},\frac{1}{\sqrt{2}}\right)=\frac{1
}{\sqrt{\pi}\sqrt{\Delta q_t^2+\Delta p_t^2 +
2\sigma(q,p)_t}} \exp
\left[-\frac{\left(X-\frac{\langle \hat{q}\rangle_t + \langle
\hat{p}\rangle_t}{\sqrt{2}} \right)^2}{\Delta q_t^2+\Delta
p_t^2 + 2\sigma(q,p)_t}\right].
\end{eqnarray}
This is  a Gaussian whose average is already determined.  Indeed,
according to the previous steps, we need two more points of
this tomograms to determine the spread $(\Delta q_t^2+\Delta
p_t^2)/2 + \sigma(q,p)_t$, from which we can retrieve
$\sigma(q,p)_t$.

Hence,  the first and second
order momenta of a Gaussian state can be measured at an arbitrary
time $t$
 by means of eight or at most ten
points belonging to three tomograms.
\section*{References}
\bibliographystyle{iopart-num}

\providecommand{\newblock}{}

\end{document}